\documentclass[aps,twocolumn,floats,superscriptaddress,prd,nofootinbib]{revtex4}
\usepackage{graphicx, epsfig, bm,amsmath}

\usepackage{color}
\usepackage{hyperref}
\usepackage{ifthen}
\usepackage{xstring}

\begin{document}


\newcommand{\zmin}{z_{\rm min}}
\newcommand{\zmax}{z_{\rm max}}
\newcommand{\lcdm}{$\Lambda$CDM}
\newcommand{\hmpc}{h^{-1}\,{\rm Mpc}}
\newcommand{\lpeak}{\ell_{\rm peak}}
\newcommand{\clpeak}{B_{\lpeak}^{2(Q)}}
\newcommand{\reff}{R_{\rm eff}}
\newcommand{\slnr}{\sigma_{\ln R}}

\newcommand{\mm}[1]{\textcolor{red}{[{\bf MM}: #1]}}
\newcommand{\wh}[1]{\textcolor{blue}{[{\bf WH}: #1]}}


\pagestyle{plain}

\title{Observational Limits on Patchy Reionization: Implications for $B$-modes}

\author{Michael J.\ Mortonson}
\affiliation{Center for Cosmology and AstroParticle Physics, 
        The Ohio State University, Columbus, OH 43210}

\author{Wayne Hu}
\affiliation{Kavli Institute for Cosmological Physics, Enrico Fermi Institute,
       and Department of Astronomy \& Astrophysics,
        University of Chicago, Chicago, IL 60637}

\begin{abstract}
The recent detection of secondary CMB anisotropy by the South Pole Telescope 
places a conservative bound on temperature fluctuations from the optical depth-modulated
Doppler effect of 
$T_{3000} < \sqrt{13}\,\mu$K at multipoles $\ell \sim 3000$.  This bound is
the first empirical constraint on reionization optical depth fluctuations at arcminute scales, 
$\tau_{3000}=0.001\,T_{3000}/\mu$K,  implying that these fluctuations are no more
than a few percent of the mean.  Optical depth modulation of the quadrupole source
to polarization generates $B$-modes that are correspondingly bounded as
 $B_{3000} = 0.003\,T_{3000}$.    The maximal extrapolation to the $\ell\sim 100$
gravitational wave regime 
yields $B_{100} = 0.1\,T_{3000}$ and remains in excess of gravitational lensing
if the effective comoving size of the ionizing regions is  $R \gtrsim 80$\,Mpc.
If patchy reionization is responsible for much of the observed 
arcminute scale temperature fluctuations,  current bounds  on $B_{100}$ already
require $R \lesssim 200$\,Mpc and can be expected to improve rapidly.
Frequency separation of thermal Sunyaev-Zel'dovich contributions to the measured secondary anisotropy would also substantially improve the limits on optical depth fluctuations and $B$-modes 
from reionization.
\end{abstract}
\maketitle


\section{Introduction}
\label{sec:introduction}

Recent observations by the South Pole Telescope (SPT) \cite{Hall:2009rv,Lueker:2009rx} and the Atacama Cosmology 
Telescope (ACT) 
\cite{ACT} 
are ushering in a new era in which our understanding of secondary anisotropy in the 
cosmic microwave background (CMB) will be revolutionized.   
Secondary anisotropy is generated after recombination by gravitational and 
scattering processes.  It is thus more
dependent on astrophysical processes than the primary anisotropy that has 
been so useful in determining fundamental cosmological parameters.
However, certain relationships between the various CMB secondary observables
can be used to scale out and, in principle, determine the unknown astrophysics. 

In this {\it Brief Report}, we discuss the example of these scaling relations provided
by patchy reionization.   Taken as an upper bound to account for contributions from other
secondary effects, SPT measurements limit optical depth fluctuations  
on arcminute scales. 
These same fluctuations generate $B$-mode polarization \cite{Hu:1999vq}
 and so the implied limits at arcminute scales
 are relatively free of both cosmological and ionization model assumptions.    
 
By making a maximal extrapolation to the degree scales relevant for
gravitational wave detection, we place
upper limits on the contamination by patchy reionization $B$-modes.  
Conversely, observational limits on degree scale $B$-modes  constrain the ionization model, in particular the size
distribution of the ionized regions, when combined with arcminute scale temperature
measurements.

\section{Scaling Relations}

Thomson scattering of CMB photons off free electrons in
linear velocity flows generates temperature fluctuations via the Doppler effect.
The first order effect from the mean optical depth during reionization is highly suppressed 
on subhorizon scales and the dominant contributions on arcminute scales
reflect optical depth modulation.  Optical depth modulations can arise from linear
density fluctuations (Ostriker-Vishniac effect), non-linear objects (kinetic Sunyaev-Zel'dovich effect), or ionization fluctuations (patchy 
reionization).

Given a measurement or an upper limit on the temperature anisotropy due to this effect, 
one can constrain optical depth fluctuations and the corresponding effect 
on $B$-mode polarization.
For notational convenience, we call the temperature power assigned to the modulated
Doppler effect at a multipole of $\ell=3000$
\begin{equation}
T_{3000}^{2(v)} \equiv {\ell (\ell+1) \over 2\pi} C_{\ell}^{TT(v)}\Big|_{\ell=3000} \,.
\end{equation}
Likewise, we use the general shorthand notation
\begin{equation}
X_{\ell}^{2(s)} \equiv
{\ell (\ell+1) \over 2\pi} C_{\ell}^{XX(s)}
\end{equation}
for optical depth ($X=\tau$) and $B$-mode
polarization ($X=B$) fluctuations; for $B$-modes, 
$s$ denotes the contribution from a particular source field.

A conservative interpretation of the SPT detection of secondary anisotropy
is that it places an upper limit of $T_{3000}^{2(v)}<13\,\mu$K$^{2}$ at 95\% CL.  This limit
assumes that all of the measured
anisotropy is assigned to the modulated Doppler effect and considers the detection as
an upper limit \cite{Hall:2009rv}.   
For reasonable cosmological models, one would expect the thermal
Sunyaev-Zel'dovich fluctuations from unresolved clusters and groups to dominate
the signal.   Thus $T_{3000}^{2(v)}\lesssim 5\,\mu$K$^2$  might serve as a more typical, albeit model dependent, limit \cite{Lueker:2009rx}.  
For this reason, we will preserve the dependence of our  main results on
$T_{3000}^{2(v)}$.

\begin{figure}[t]
\centerline{\psfig{file=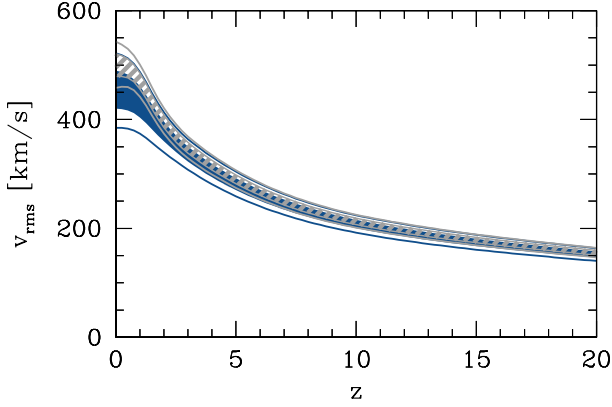, width=3.3in}}
\caption{Predicted range of $v_{\rm rms}(z)$ 
(shading: 68\% CL region; curves: 95\% CL region)
for flat \lcdm\ (light gray) and non-flat quintessence (dark blue) 
models constrained by current data.
}
\label{fig:vrms}
\end{figure}

Optical depth fluctuations with a power spectrum $C_{\ell}^{\tau\tau}$ modulate the
Doppler effect from a velocity field with rms $v_{\rm rms}$ to produce
a temperature power spectrum \cite{Hu:1999vq}
\begin{equation}
C_\ell^{TT(v)} \approx {1 \over 3} C_\ell^{\tau\tau} v_{\rm rms}^2
\label{eq:ttauscaling}
\end{equation}
below the coherence scale of the flows.
The factor of 3 here accounts for the line-of-sight nature of the Doppler
effect.  Note that for simplicity we
assume optically thin conditions throughout and ignore the 10--20\% effects from
the finite mean opacity during reionization.  Given the linear theory matter 
power spectrum  today, $P_{\rm lin}(k)$,
the rms velocity is computed as
\begin{equation}
v_{\rm rms}^2(z) = \left[\frac{H(z)}{1+z}\,{ d D_1 \over d\ln a} \right]^2 
\int \frac{dk}{2\pi^2} P_{\rm lin}(k),
\end{equation}
where $D_1$ is the linear growth function for density fluctuations.
Figure~\ref{fig:vrms} shows the range of $v_{\rm rms}(z)$ allowed by 
current constraints on flat \lcdm\ models and non-flat quintessence 
models with arbitrary variations in the dark energy equation of state at $z<1.7$,
including measurements of 
the CMB, supernovae, baryon 
acoustic oscillations, and the Hubble constant as described in 
Ref.~\cite{PCcurrent}.   Note that predictions for $v_{\rm rms}$ at $z > 2$
have little uncertainty from cosmological parameters including variations in 
curvature and dark energy, with the possible exception of the presence of
a substantial component of the energy density in dark energy or massive neutrinos
at high redshift.

The same optical depth fluctuations modulate the generation of 
polarization from the primordial quadrupole with rms $Q_{\rm rms}$.  In the 
Sachs-Wolfe approximation \cite{Hu:1999vq},
\begin{equation}
Q_{\rm rms}^2(z) \approx \frac{1}{60} A_s(k_0) [k_0(\eta(z)-\eta_*)]^{1-n_s}
\Gamma_{\rm SW}(n_s)\,,
\end{equation}
where the initial curvature spectrum normalized at the scale $k_0$ is 
$\Delta_{\cal R}^2(k) = A_s (k/k_0)^{n_s}$, 
$\eta$ is conformal time with 
$\eta_*$ evaluated at recombination,
and
\begin{equation}
\Gamma_{\rm SW}(n_s) = 3 \sqrt{\pi}~{\Gamma[(3-n_s)/2] \Gamma[(3+n_s)/2]
\over \Gamma[(4-n_s)/2]\Gamma[(9-n_s)/2] }\,.
\end{equation}
At plausible redshifts for reionization, $z\sim 10$, the predicted quadrupole 
for flat \lcdm\ models is a nearly constant
$Q_{\rm rms}=18.2\pm 0.5~\mu$K, 
using the same data as for the $v_{\rm rms}$ 
predictions in Fig.~\ref{fig:vrms}. The uncertainty in $Q_{\rm rms}$ is 
negligible compared with $v_{\rm rms}$.

  Modulation destroys the
symmetry that produces only $E$-modes from scalar perturbations, 
generating equal power in $E$ and $B$ polarization. When scaled to the 
temperature spectrum, the $B$-mode power spectrum from the modulated quadrupole becomes
\cite{Hu:1999vq}
\begin{equation}
C_\ell^{BB(Q)} \approx {9 \over 100}\left({  Q_{\rm rms} \over v_{\rm rms}}\right)^2  C_\ell^{TT(v)} \,.
\label{eq:btscaling}
\end{equation}
Note that this scaling relation remains true regardless of whether the optical
depth fluctuations are due to density effects like the Ostriker-Vishniac anisotropy
\cite{Vis87} or
ionization effects from inhomogeneous reionization.
The only difference between these effects is the effective redshift at which
$Q_{\rm rms}/v_{\rm rms}$ is evaluated, with $v_{\rm rms}$ dominating the variations.    To maintain generality, we keep
both $Q_{\rm rms}$ and $v_{\rm rms}$ in the relations but scale their values
to $z \sim 10$. 

Modulation of the $e^{-\tau}$ screening of the primary $E$-modes also 
generates $B$-modes below the coherence scale of the $E$-modes, i.e.~the $\ell \sim 10^{3}$ damping scale of the primary anisotropy.  
On these scales, the screening $B$-modes take the form 
\cite{Dvorkin:2009ah}
\begin{equation}
C_\ell^{BB(E)} \approx {3 \over 2}\left({  E_{\rm rms} \over v_{\rm rms}}\right)^2  C_\ell^{TT(v)}\,,
\end{equation}
where 
\begin{equation}
E_{\rm rms}^{2} \approx \sum_{\ell} {2\ell+1 \over 4\pi}C_{\ell}^{EE} \,.
\end{equation}
$E_{\rm rms}=6.4\,\mu$K for the maximum likelihood $\Lambda$CDM model 
and varies little with cosmological parameters.

From Eq.~(\ref{eq:ttauscaling}), optical depth fluctuations are related to the temperature anisotropy $T_{3000}^{(v)}$ as
\begin{equation}
\tau_{3000} \approx
0.00095  \, {T_{3000}^{(v)} \over \mu{\rm K}}\,
{200\,{\rm km/s} \over v_{\rm rms}}\,.
\end{equation}
Hence for the strict upper limit of $T_{3000}^{(v)}<\sqrt{13}\,\mu$K, the rms fluctuation in $\tau$
at a few arcminutes is $\tau_{3000} < 0.003$, i.e. 
no more than a few percent of the mean optical depth 
(e.g. $\bar \tau=0.10\pm 0.02$ for general reionization histories at $6<z<30$
constrained by 5-year WMAP data \cite{WMAP5reion}).
Models that predict a percent optical depth rms or $\sim 10\%$ of the
mean as a typical fluctuation are observationally unviable (cf.~\cite{HolIllMel07}).

For $B$-modes from the modulated quadrupole [Eq.~(\ref{eq:btscaling})],
\begin{equation}
B_{3000}^{(Q)}
\approx
0.003 \, T_{3000}^{(v)} \, {Q_{\rm rms}\over 18\,\mu{\rm K}}\,
 {200\,{\rm km/s} \over v_{\rm rms} }
 \,.
\end{equation}
Combining quadrupole and $E$-mode screening modulation yields
\begin{eqnarray}
B_{3000}^{2(Q+E)} & \approx   &
 \left( 0.003 \, T_{3000}^{(v)} \, \frac{200\,{\rm km/s}}{v_{\rm rms}}\right)^{2} \\
&& \times \left[ \left( {Q_{\rm rms}\over 18\,\mu{\rm K}}\right)^2 
+ 2.1 \left( {E_{\rm rms}\over 6.4\,\mu{\rm K}}\right)^2 \right]  
 \nonumber
\end{eqnarray}
as the total $B$-modes from reionization at $\ell = 3000$.
Thus upper limits on $T_{3000}^{(v)}$ place stringent bounds on patchy reionization $B$-modes
given predictions for $v_{\rm rms}$, $Q_{\rm rms}$, and $E_{\rm rms}$.
For the strict upper limit $T_{3000}^{(v)}<\sqrt{13}\,\mu$K, 
$B_{3000}^{(Q)}<0.01\,\mu$K.

\begin{figure}[t]
\centerline{\psfig{file=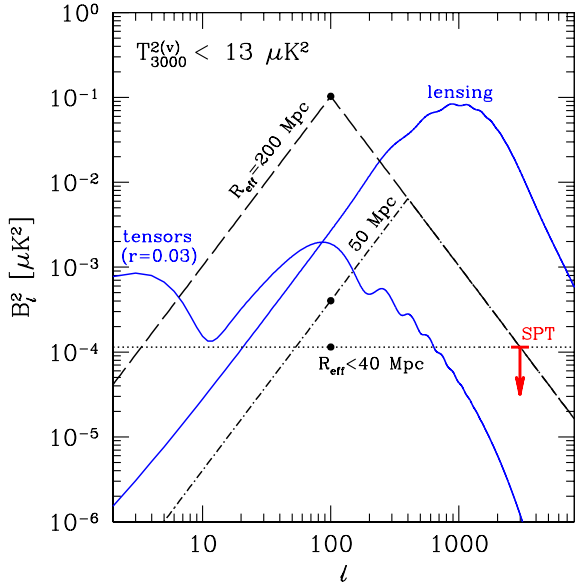, width=3.3in}}
\caption{Extrapolation of the inferred SPT limit on modulated quadrupole
$B$-modes (red arrow at $\ell=3000$ with $T_{3000}^{2(v)} < 13\,\mu {\rm K}^2$, $z\approx 10$) 
to limits at $\ell\sim 100$ 
(black dots) using the sharp peak model with a single bubble size of 
$R_{\rm eff}=200$~Mpc (black dashed lines) or $R_{\rm eff}=50$~Mpc 
(black dot-dashed lines).  For $R_{\rm eff} \lesssim 40$~Mpc, 
we take a more conservative upper limit shown
by the horizontal dotted line. For reference, solid blue curves show the 
contributions to $B_{\ell}^2$ expected from lensing and gravitational 
waves with tensor-to-scalar ratio $r=0.03$
for the best fit flat \lcdm\ model.
}
\label{fig:bl}
\end{figure}

\section{Large angle $B$-mode limits}

To relate the $B$-mode contributions at $\ell=3000$ to the 
large angle regime at $\ell \sim 100$
relevant for gravitational wave studies we require a model for the optical depth fluctuations.
Note that the screening contributions die off as white noise for $\ell < \ell_{A} \sim 300$ given the
acoustic scale and cannot generate large contributions at $\ell \sim 100$
\cite{Dvorkin:2009ah}.  We thus need to relate the modulated quadrupole contributions
between the two scales.

To place upper limits on the $B$-mode contribution we begin by assuming all
of the temperature power is due to ionization fluctuations 
since the $B$-mode signal from density fluctuations is expected 
to be well below that of gravitational lensing 
at all scales \cite{Mortonson:2006re}.  
For a variety of simple analytic models of patchy reionization 
that assume completely ionized, spherical bubbles with a lognormal 
distribution of bubble radii \cite{ZalFurHer04,WanHu05}, the 
modulated quadrupole contribution to $B$-modes can be approximated as a single 
peak in $B_{\ell}^{2}$ that rises as $\ell^2$ at 
$\ell \ll \lpeak$ and falls as $\ell^{-2}$ at $\ell \gg \lpeak$, where
$\lpeak \approx 2\times 10^4 (\reff/{\rm Mpc})^{-1}$ \cite{Mortonson:2006re}.
The effective ionized bubble radius $\reff$ is determined by both 
the volume-averaged radius of bubbles $R_V$ and the lognormal width of the 
bubble size distribution $\slnr$: $\reff=R_V \exp(2.5\slnr^2)$.

\begin{figure}[t]
\centerline{\psfig{file=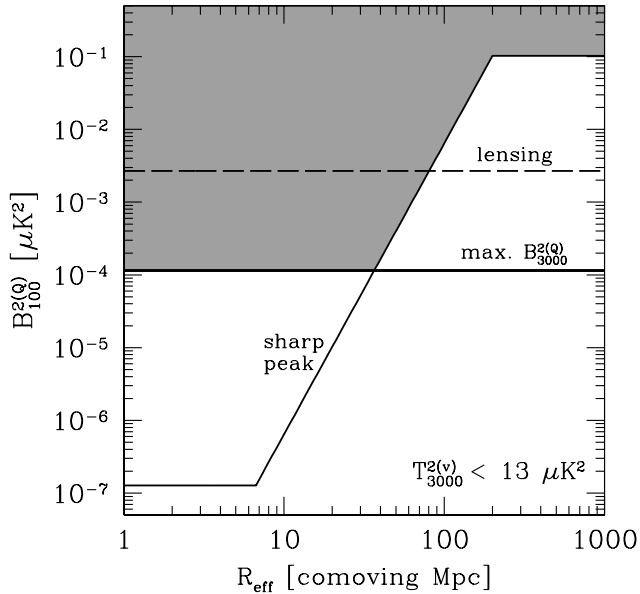, width=3.3in}}
\caption{Upper limits 
on the $B$-mode power from patchy reionization at $\ell=100$
as a function of the effective radius of ionized bubbles, 
using $T_{3000}^{2(v)}<13\,\mu{\rm K}^2$ and assuming $z \approx 10$.
The thin solid line extrapolates from the upper limit on $B_{3000}^{2(Q)}$ to 
$\ell=100$ using 
the scaling of Eq.~(\ref{eq:sharppeak}), and the thick solid line shows 
the extrapolation assuming a flat spectrum. 
The shaded region above both of these curves is excluded.
The $B$-mode power at $\ell=100$ from lensing in the best fit 
flat \lcdm\ model is shown as a dashed line.
}
\label{fig:rbound}
\end{figure}

Let us start by taking a single bubble size, i.e. a delta function in the
distribution with 
 $\slnr\to 0$.   
Then the shape of the patchy reionization signal, shown in Fig.~\ref{fig:bl},
is approximately described as a 
sharp peak\footnote{Ringing in $\ell$ 
produced by a delta function distribution is smoothed out by 
any more realistic
bubble distribution and by the width in redshift of the 
reionization transition.}
\begin{equation}
B_{\ell}^{2(Q)}\approx \left\{
\begin{array}{l}
\left(\frac{\ell}{\lpeak}\right)^2 \clpeak\,, \,\, \ell\le\lpeak\,, \\
\left(\frac{\ell}{\lpeak}\right)^{-2} \clpeak\,, \,\, \ell>\lpeak\,.
\end{array}\right.
\end{equation}
The $B$-mode  rms at $\ell=100$ scales with  $R_{\rm eff}$
as 
\begin{equation} 
B_{100}^{(Q)}
\approx
2.2 \times 10^{-6}  \, T_{3000}^{(v)}\, {Q_{\rm rms}\over 18\,\mu{\rm K}}\,
 {200\,{\rm km/s} \over v_{\rm rms} } \left(  {R_{\rm eff} \over{\rm Mpc}} \right)^2
 \,
\label{eq:sharppeak}
\end{equation}
for $7 \lesssim  R_{\rm eff}/{\rm Mpc} \lesssim 200$ and saturates outside this
range, as shown in Fig.~\ref{fig:rbound}.

This assumption of a sharply peaked spectrum provides a conservative 
extrapolation of upper limits from $\ell\sim 3000$ to $\ell\sim 100$ if 
$\lpeak \lesssim 500$ (or $R_{\rm eff}\gtrsim 40$ Mpc).
In this case, realistic bubble size distributions
would produce a flatter spectrum and thus
less power at $\ell = 100$ (see \cite{Mortonson:2006re}, Fig. 7).  

In the opposite limit of $R_{\rm eff}\lesssim 40$ Mpc, a flatter distribution would produce
more power at $\ell\sim 100$ than a sharp peak and so we conservatively assume $B_{100}^{(Q)}\geq B_{3000}^{(Q)}$.
Using this model to extrapolate from the SPT limit at $\ell\sim 3000$
to $\ell\sim 100$ produces an upper bound shown in Fig.~\ref{fig:rbound} as the shaded
excluded region.

Note that as the bubble size increases, the upper limit at $\ell=100$ saturates.  This
saturation is essentially model independent and corresponds to a maximal slope of 
$\ell^{-2}$ connecting the power at $\ell=100$ and $3000$.  
This strict upper bound is given by 
\begin{equation}
B_{100}^{(Q)}
< 0.09\, T_{3000}^{(v)}\, {Q_{\rm rms}\over 18\,\mu{\rm K}}\,
{ 200\,{\rm km/s} \over v_{\rm rms} }    \,.
\end{equation}
With $T_{3000}^{(v)}<\sqrt{13}\,\mu$K, this limit still allows more power than the lensing $B$-modes, $B_{100}^{2(L)} \approx (2.68 \pm 0.25)\times 10^{-3}\,\mu{\rm K}^2$ 
(with fractional errors approximately scaled as $25 \sigma(\Omega_c h^2)$ \cite{Smith:2006nk}).
 
Conversely, observational limits on $B_{100}$ can be combined with measurements of
 $T_{3000}^{(v)}$ to constrain the bubble size distribution. 
If the distribution is dominated by a single 
effective comoving radius, Eq.~(\ref{eq:sharppeak}) provides an estimate of 
the required precision to place an upper bound on this radius of 
\begin{equation}
R_{\rm eff} \sim 200 ( B_{100}^{(Q)} / 0.1\,T_{3000}^{(v)} )^{1/2}\,{\rm Mpc}
\end{equation}
Current bounds of 
$B_{100}^2 < 0.1\,\mu$K$^2$ (95\% CL) from BICEP  \cite{Chiang:2009xsa}
in fact exclude $R_{\rm eff} \gtrsim 200\,$Mpc  for the maximal
$T_{3000}^{(v)}= \sqrt{13}\,\mu$K.

\section{Discussion}

We have shown that the SPT detection of secondary temperature anisotropy interpreted as a limit
on modulated Doppler contributions from scattering of $<\sqrt{13}\,\mu$K provide the first empirical
bounds on optical depth fluctuations during reionization at a scale of a few arcminutes.   
Models that produce fluctuations on these scales in excess of a few percent of the mean
are no longer viable.
This limit on optical depth variations in turn produces a nearly model- and cosmology-independent 
limit on the $B$-mode polarization from reionization of $< 0.01\, \mu$K 
 at similar scales.  
We expect both limits to improve dramatically once more frequency channels of the data 
and larger regions of the sky have been analyzed so that the presumably dominant 
thermal Sunyaev-Zel'dovich contribution can be better separated.  
 
 Such upper limits can be turned into constraints on degree scale $B$-mode
 contamination of the 
 gravitational wave signal.   The maximal allowed $B$-modes
 at $\ell \approx 100$ can still exceed those from gravitational lensing, but only if the effective
 radius of the ionized regions is greater than $\sim 80$\,Mpc.
In fact, current direct limits on degree scale $B$-mode polarization require 
ionized regions
$\lesssim 200$\,Mpc  for the maximal allowed fluctuations at arcminute scales.    

Hence expected improvements in both small angle temperature measurements and
large angle $B$-mode polarization measurements  should rapidly advance our 
empirical understanding of reionization.

\medskip {\it Acknowledgments:} 
 We thank Tom Crawford for useful conversations.  
MJM was supported by CCAPP at Ohio State.
WH was supported by the KICP under
NSF contract PHY-0114422, DOE contract DE-FG02-90ER-40560 and the Packard Foundation.

\vfill
\bibliographystyle{arxiv_physrev}
\bibliography{sptbounds}

\end{document}